\documentclass[a4paper,draft,aps,prl,preprint,superscriptaddress,nofootinbib]{revtex4}
\usepackage[latin1]{inputenc}
\usepackage{amsmath}

\usepackage{url}
\usepackage{amsfonts}
\usepackage{amssymb}
\usepackage[final]{graphicx}
\usepackage[normalem]{ulem}
\usepackage[usenames, dvipsnames]{color}
\usepackage{soul}
\usepackage{color}

\begin{document}
\author{Zhiyue Lu}
\email{zhiyuelu@gmail.com}
\affiliation{James Franck Institute, University of Chicago, IL 60637, U.S.A.}
\author{Oren Raz}
\email{orenraz@gmail.com}
\affiliation{Department of Chemistry and Biochemistry ,
	University of Maryland, College Park, MD 20742, U.S.A.}
\title{Anomalous cooling and heating -- the Mpemba effect and its inverse}
\begin{abstract}
Under certain conditions, it takes a shorter time to cool a hot system than to cool the same system initiated at a lower temperature. This phenomenon -- the ``Mpemba Effect'' -- is well known in water, and has recently been observed in other systems as well. However, there is  no single generic mechanism that explains this counter-intuitive behavior. Using the theoretical framework of non-equilibrium thermodynamics, we present a widely applicable mechanism for this effect,  derive a sufficient condition for its appearance in Markovian dynamics, and predict an inverse Mpemba effect in heating:  under proper conditions, a cold system can heat up faster than the same system initiated at a higher temperature. Our results suggest that it should be possible to observe the Mpemba effect and its inverse in a variety of systems, where they have never been demonstrated before. 
\end{abstract}
\maketitle
\paragraph{Introduction.}
		Consider two cups of water, prepared at different initial temperatures, $T_h>T_c$, but identical in all other macroscopic parameters. When coupled to a cold bath with temperature $T_b$ (where $T_b<T_{c}$), which one takes a shorter time to freeze?  Surprisingly, under certain conditions the hot water freezes first. This observation is known as the ``Mpemba effect'' \cite{AmJPhys_General_jeng2006mpemba}, and various properties of water have been considered to explain it: supercooling \cite{SupperCooling_auerbach1995supercooling,Experiment_esposito2008mpemba}; evaporation \cite{Theory_Evap_Colling_vynnycky2010evaporative,Theory_evaporation_kell1968freezing}; convection flow \cite{Theory_NotOnlyConv_vynnycky2015can}; differences in dissolved gases and solids \cite{Exp_GassesEff_wojciechowski1988freezing,Disolved_katz2009hot}; and the anomalous relaxation of the hydrogen bond \cite{Theory_HydroBond_zhang2014hydrogen,Molecular_Dynamics_jin2015mechanisms}. 
		 In recent years similar behavior has been observed in other substances, e.g. carbon nano-tubes resonators \cite{Theory_Carbon_nano_greaney2011mpemba} and magneto-resistance alloys \cite{Magnetic_banerjee2011history}. These observations indicate that such an anomalous cooling effect is generic and not exclusive to water. 
		 
		Anomalous cooling is counter-intuitive because it differs from quasi-static cooling (i.e. Newton's heat law), where the system's temperature gradually decreases toward the bath's temperature, thus the hot system first cools to the initial temperature of the cold system and must lag behind it. However, cooling by interacting with a cold bath (i.e. quenching) is in general not quasi-static, but rather a genuinely far-from-equilibrium process. The Mpemba effect is possible when the hot system takes a non-equilibrium ``shortcut'' in the system's state-space and bypasses the cold system. Is there a general non-equilibrium mechanism underlying all anomalous cooling effects?

		Here we study the Mpemba effect with the recently developed tools of stochastic thermodynamics \cite{seifert_review_2012}. We first present a  physical intuition and  provide a numerical example for the effect in a 1-d potential. 
		We then analytically study the effect in Markovian dynamics, and provide the exact mathematical conditions for its occurrence. In addition, we predict an {\it inverse Mpemba effect} -- under certain conditions, an initially cold system heats up faster than the same system initiated at a warmer temperature, when both are heated by the same hot bath. This effect has not been reported experimentally.
		We stress that the mechanism suggested here does not intend to replace any existing explanations of the Mpemba effect in water. Rather, it is a novel approach to describe and predict anomalous cooling and heating effects in a variety of systems, which can be experimentally tested.

		For clarity, let us first formulate a generic description of Mpemba effect in an arbitrary system. Consider the following experiment: two identical copies of the system are prepared at different temperatures $T_h>T_c$, and are simultaneously cooled by a very cold bath with temperature $T_b$ ($T_b<T_c$). Their cooling process is tracked and characterized by the decay of their {\it distance from equilibrium} function (defined later). The distance from equilibrium of the hot system is initially larger than that of the cold system. If there is some critical time $t_m$ such that for any $t>t_m$ the distance from equilibrium of the initially hot system is smaller than that of the the initially cold system, then the Mpemba effect occurs. 
		
\begin{figure}[h]
			\centering 				 
			\includegraphics[scale=0.55]{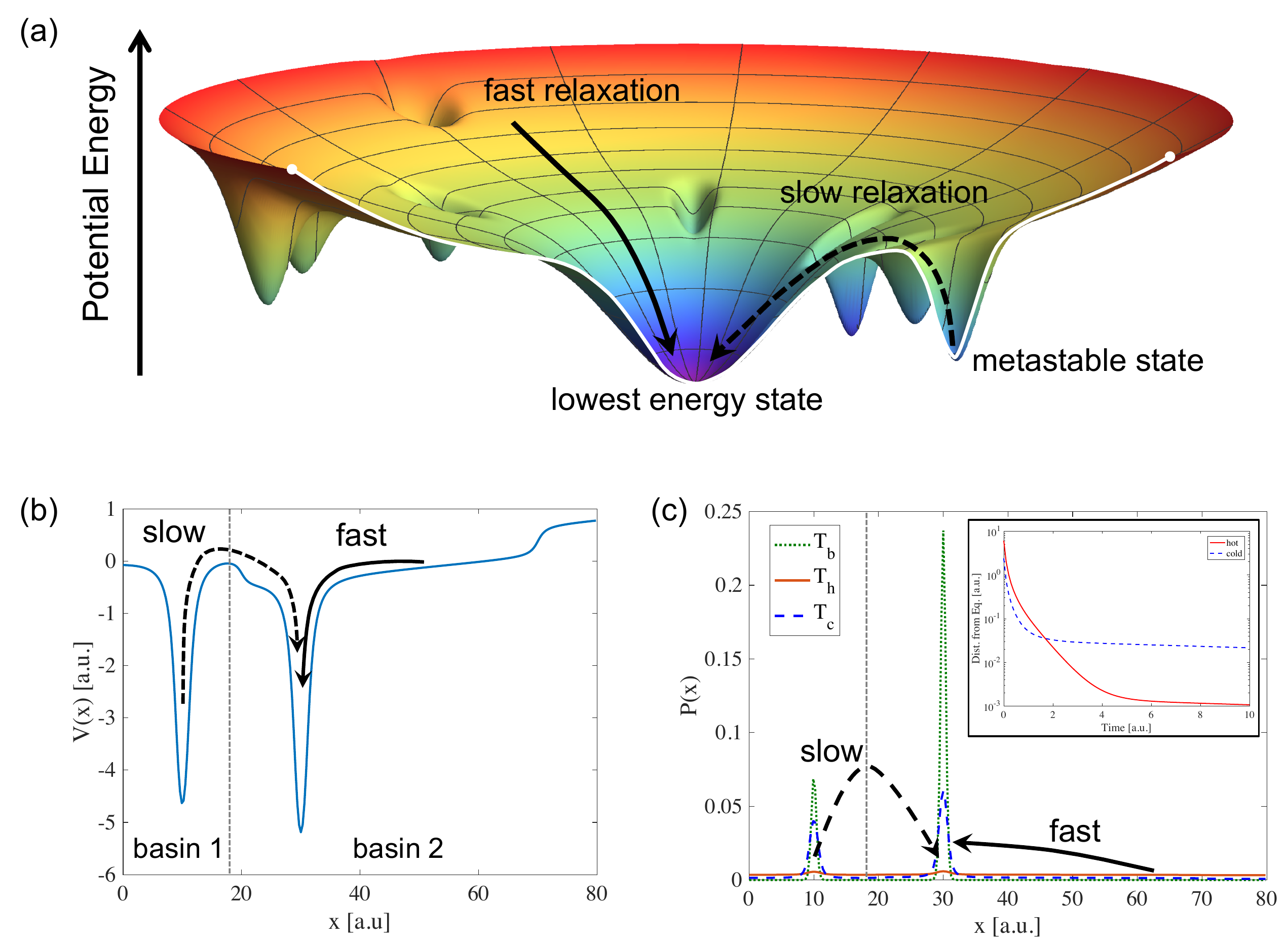}\caption{(a) For illustration purpose, we schematically represent a high dimensional configuration space by a 2-dimension manifold, that is sketched as a funnel-shaped energy landscape \cite{Link}. The funnel is cut open for clarity. The lowest energy state is achieved at the origin. The solid arrow represents a fast relaxation and the dashed arrow represents a slow relaxation. (b) As an example, we construct a 1-d energy landscape and demonstrate the Mpemba effect in the corresponding Fokker-Planck dynamics. The well on the left is a meta-stable state, whereas the well on the right represents the lowest energy state. Note that the basin width of the deeper well is larger than that of the shallow well. (c) The Boltzmann distributions at different temperatures. After the quench, both initial distributions relax toward the final equilibrium distribution (dotted green). Although the initially colder (dashed blue) system is more populated in the lowest well compared with the initially hot system (solid red), after a short time of relaxation, the hot system ends up with higher population in the lowest well due to the fast relaxation from its basin. This grants the initially hot system an advantage over the colder one, and the Mpemba effect occurs. }
			\label{fig:Fig_1_landscape}
		\end{figure}

		\paragraph{Energy Landscape and the Mpemba effect:}
		The systems that were shown to have the Mpemba effect so far share very little in common, but they all have complicated potential energy landscapes dictating their dynamics. In what follows, we argue heuristically how certain geometric features of the energy landscape can lead to the Mpemba effect. As an example, we numerically demonstrate this argument in a 1-d diffusion process on the energy landscape shown in Fig.~(\ref{fig:Fig_1_landscape}b). An exact mathematical analysis follows.

		As schematically illustrated in Fig.~(\ref{fig:Fig_1_landscape}a), a rough energy landscape of a thermodynamic system commonly consists of multiple energy wells separated by energy barriers. When the system interacts with a thermal bath, it eventually relaxes into a unique equilibrium distribution. At short timescales, these relaxations are \emph{localized} within basins of the energy landscape, and at longer timescales the relaxation can transport probability between basins ({\it inter-basin} relaxation). Integrating the probability distribution within each basin, we define a {\it coarse-grained distribution}, which evolves slowly by the inter-basin relaxation.		
		
		In the Mpemba experiment, two systems are prepared at the Boltzmann's distributions $\pi(T_h)$ and $\pi(T_c)$ corresponding to the  temperatures $T_h$ and $T_c$. They are both different from the final equilibrium distribution $\pi(T_b)$ at the bath's temperature $T_b$. Nevertheless, it is possible that the coarse-grained distributions of $\pi(T_b)$ and $\pi(T_h)$ are almost identical, but the coarse-grained distribution of $\pi(T_b)$ and $\pi(T_c)$ are quite different. In this situation, the initially hot system rapidly relaxes toward the final equilibrium through the fast localized relaxation, whereas the initially cold system relaxes slowly due to the inter-basin relaxation. As a result cooling the hot system takes shorter time than cooling the  cold one.	
				
		This argument is demonstrated by an example of a diffusion process (the Fokker-Planck equation) in the 1-d potential landscape shown in Fig. (\ref{fig:Fig_1_landscape}b).  At high initial temperature $T_h$ the energy landscape plays only a minor role and the initial distribution is spread almost uniformly over the configuration space. Therefore, the integrated coarse-grained probability of each basin is proportional to its width. The widths of the two basins were chosen to match the coarse-grained distribution of $\pi(T_b)$ (see Fig.~(\ref{fig:Fig_1_landscape}c)). Thus, the relaxation from the initially hot system involves mainly the fast relaxations. In contrast, the coarse-grained distribution of the initially cold system $\pi(T_c)$ is very different from that of the final equilibrium and its relaxation involves both the localized and the slower inter-basin relaxations. As a result, the initially hot system cools faster than the initially cold one. The distances from equilibrium (defined later on) are plotted in Fig.~(\ref{fig:Fig_1_landscape}c). See SI for a detailed discussion of the example.

		\paragraph{Markovian Dynamics:}
		So far we have provided a qualitative picture of the Mpemba effect. To quantitatively study the effect we next describe the evolution of the system by a continuous-time Markovian dynamic. To ensure that the system relaxes into a unique and genuine equilibrium, we demand that the dynamic is ergodic and satisfies detailed balance. To simplify the presentation, we only discuss systems with finite configuration space\footnote{However, note that a similar analysis can be carried out for continuous operators as well, e.g. for the Fokker-Planck operator of the system in Fig. (\ref{fig:Fig_1_landscape}b).}. At time $t$, the probability to find the system at a certain state $i$ is denoted by $p_i(t)$, and we can characterize the system by the \emph{probability distribution vector}, $\vec p(t) = \left(p_1(t),...,p_n(t)\right)$. Due to thermal fluctuations, a system in contact with a heat bath evolves stochastically and performs random transitions between its different states. The rate of transition from state $j$ to state $i$ ($R_{ij}$), is jointly determined by the energy ($E_j$), the energy barrier between the states ($B_{ij}$), and the temperature of the thermal bath ($T_b$). More precisely, the system coupled to the bath $T_b$ evolves according to the master equation:
		\begin{eqnarray}\label{Eq:Master}
		\dot p_i(t) = \sum_j R_{ij}p_j(t)~~\text{for $i=1,2,\cdots,n$}.
		\end{eqnarray} 
		  The elements $R_{ij}$ of any detailed balanced transition rate matrix can be written as \cite{mandal2011proof} 
		\begin{eqnarray}\label{Eq:R_def}
		R_{ij} = \begin{cases}
		\Gamma e^{-\frac{B_{ij}-E_j}{k_B T_{b}}} & i\neq j\\
		-\sum_{k\neq i} R_{ki} & i=j
		\end{cases}
		\end{eqnarray}
		where $\Gamma$ is a constant rate that fixes the dimensions, $B_{ij}=B_{ji}$ and the diagonal terms $R_{ii}$ are defined such that  the normalization, $\sum_i p_i(t) = 1$, is conserved. Under this dynamic, any initial distribution eventually relaxes into the equilibrium (Boltzmann) distribution
		\begin{equation}\label{Eq:Boltzmann}
		\pi_i( T_b) = \frac{e^{-E_i/k_B T_{b}}}{\sum_i e^{-E_i/k_B T_{b}}}.
		\end{equation}

		\begin{figure}[htbp]
			\centering
			\includegraphics[width=\linewidth]{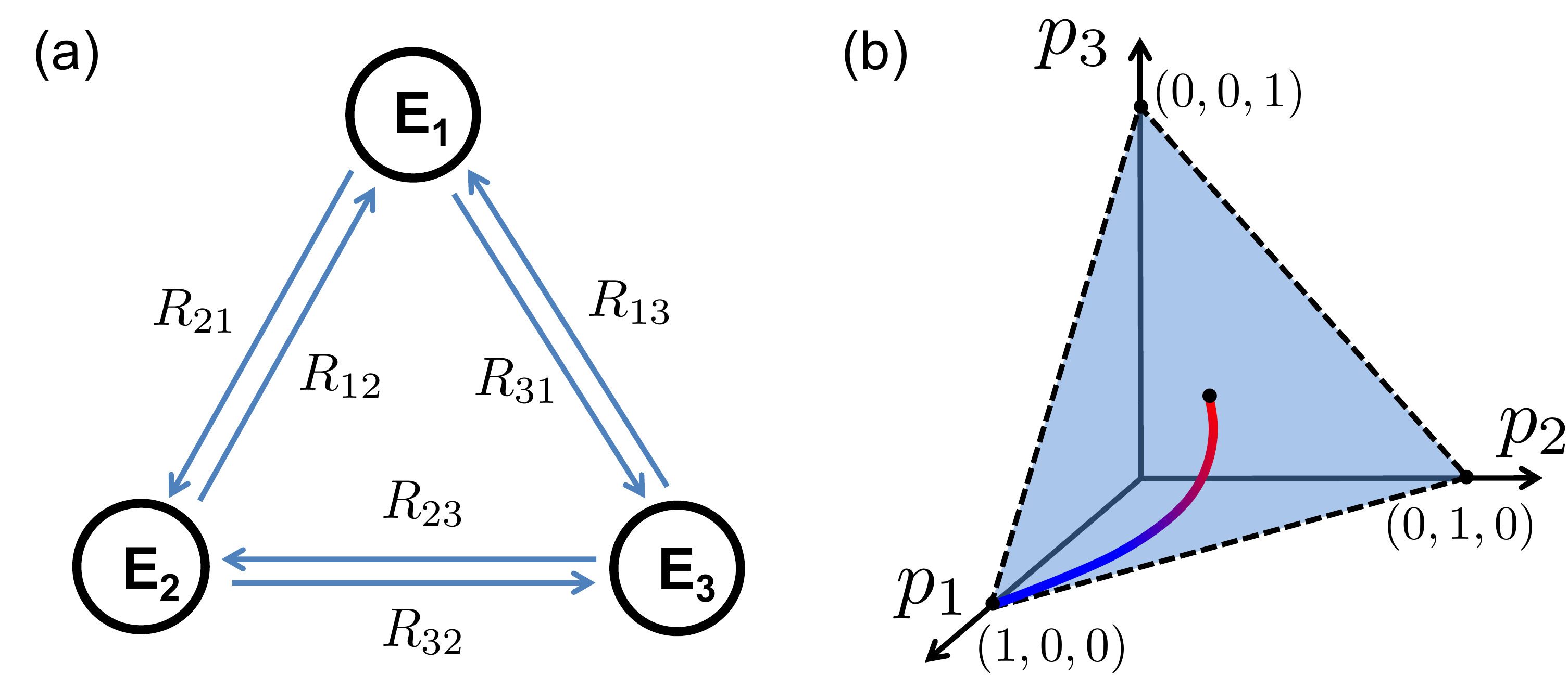}
			\caption{ (a) A  three-state system with Markovian dynamics, where $R_{ij}$ denotes transition rate from $j$ to $i$. (b) The probability distribution among the three states can be described by the vector $\vec p=(p_1,p_2,p_3)$, and all possible values of $\vec p$ forms a simplex (shaded triangle) in the $(p_1,p_2,p_3)$ space. The curved line in this simplex is the quasi-static locus, i.e. the set of Boltzmann distributions $\vec \pi(T)$ corresponding to temperatures from $0$ to $\infty$. }
			\label{Fig:model}
		\end{figure}
		
		\paragraph{The distance function} In the original discussion on the Mpemba effect, the cooling rate was characterized by measuring the time it takes for the water to freeze\footnote{this is a subtle definition, see discussion in \cite{AmJPhys_General_jeng2006mpemba}.}. In our analysis we do not rely on phase transitions, rather, we quantify the rate of cooling by constructing a distance-from-equilibrium function, $D[\vec p(t);T_b]$, and observing its decay over time. $D[\vec p;T_b]$ measures the distance of any distribution $\vec p$ from the  equilibrium Boltzmann distribution $\vec \pi(T_b)$ given in Eq.(\ref{Eq:Boltzmann}). The Mpemba effect occurs if there exist a $t_m$ such that $D[\vec p^h(t); T_b]<D[\vec p^c(t); T_b]$ for all $t>t_m$, provided that at $t=0$, the initially hot system had a larger distance from equilibrium,  $D[\vec p^h(0); T_b]>D[\vec p^c(0); T_b]$. 
		
		There are many reasonable choices for a distance-from-equilibrium function. However, the Mpemba effect may be falsely reported due to a poor choice of the distance function $D[\vec p;T_b]$. To avoid such cases, we demand that $D[\vec p;T_b]$ satisfies the following three properties: (i) When the system relaxes towards its thermal equilibrium (from any initial distribution), the value of $D[\vec p(t);T_b]$ monotonically non-increasing with time;  (ii) The distance from equilibrium of a system initiated at temperature $T_h$ is larger than that of the same system initiated at $T_c$ when $T_b<T_c<T_h$. In other words, $D[\vec \pi (T);T_b]$ is a monotonically increasing function of $T$ for $T>T_b$.  Lastly, (iii) the distance from equilibrium is a continuous, convex function of $\vec p$. As we show below, the identification of the  Mpemba effect is indifferent to the specific choice of the distance function, as long as these three conditions are satisfied. In what follows\footnote{Alternative distance functions that satisfy all the three requirements are the Kullback-Leibler divergence and the $L_1$ distance between $\vec p$ and $\vec \pi (T_b)$. } we use the {\it entropic distance} between any distribution $\vec p(t)$ and the equilibrium distribution $\vec\pi(T_b)$, defined by the total amount of entropy produced in the relaxation from $\vec p$ to $\vec \pi(T_b)$ (See SI):
		\begin{align}\label{Eq:DistFromEq}
		D_e[\vec p(t);T_b] =\sum_{i}\left(\frac{E_i \Delta p_i}{T_b}  +p_i\ln p_i-\pi^b_i\ln\pi^b_i\right), 
		\end{align}
		where $\pi^b_i$ denotes the equilibrium probability at temperature $T_b$, $\Delta p_i(t)=p_i(t)-\pi^b_i$, and the units are set such that $k_B=1$.  $D_e[\vec p(t);T_b]$ is a continuous convex function of $\vec p$, and by the second law of thermodynamics $D_e[\vec p(t);T_b]$ is a monotonically decreasing function of time (see \cite{seifert_review_2012} and SI for a proof). Moreover, as we show in the SI, $D_e[\vec \pi (T);T_b]$ is a monotonic function of $T$ for $T>T_b$. Therefore, this choice satisfies all the three requirements for $D[\vec p,T_b]$.

		\paragraph{A sufficient condition for the Mpemba effect.} 
		To find the conditions under which the Mpemba effect occurs, it is useful to study the relaxation trajectories $\vec p^h(t)$ and $\vec p^c (t)$ in probability space. These can be analyzed using the ordered eigenvalues $\lambda_1>\lambda_2\geq \lambda_3 \geq ... \geq \lambda_n$ and corresponding right eigenvectors $\vec v_1,\vec v_2,...,\vec v_n$ of the rate matrix $R_{ij}$. Since $R_{ij}$ satisfies detailed balance, it is diagonalizable with all the eigenvalues real \cite{schnakenberg1976network,kube2006coarse}, and $0=\lambda_1>\lambda_2$. The null eigenvector $\vec v_1$ equals the unique equilibrium distribution $\vec\pi( T_b)$. We can express any initial probability distribution as a linear combination of the eigenvectors, $\vec p (0) = \vec \pi( T_b) + a_2\vec v_2 +...+a_n \vec v_n$. In addition, the probability evolves in time according to 
		\begin{eqnarray}
		\vec p(t) = \vec\pi( T_b) + e^{\lambda_2 t}a_2 \vec v_2 +...+ e^{\lambda_n t}a_n \vec v_n .
		\end{eqnarray}
		$\lambda_2$  is the slowest relaxation rate in the system. When  $\lambda_2>\lambda_3$ and $\left|a_2^c\right|>\left|a_2^h\right|$,  the Mpemba effect occurs. This can be seen from the following argument: for large enough $t$,  the terms $e^{\lambda_k t}a_k \vec v^k$ for $k\geq 3$ are exponentially smaller than $e^{\lambda_2 t}a_2 \vec v_2$. Therefore, they can be neglected, and the two relaxation trajectories can be written as $\vec p^h(t) \approx \vec\pi(T_b) + a_2^h \vec{v}_2 e^{\lambda_2t}$ and $\vec p^c(t) \approx \vec\pi(T_b) + a_2^c \vec{v}_2 e^{\lambda_2t}$. Therefore, at large time the relaxation trajectories $\vec p^c(t)$ and $\vec p^h(t)$ follow almost the same trajectory, and since $|a_2^c|>|a_2^h|$,  $\vec p^c(t)$ lags behind $\vec p^h(t)$. By the convexity and monotonicity of $D[\vec p(t); T_b]$ this implies the Mpemba effect as defined above (see SI for a proof).   
		
		To identify whether a system can show the Mpemba effect at a given $T_b$, we next consider the \emph{quasi-static locus} $\vec \pi(T)$ \cite{callen2006thermodynamics}, which is the set of Boltzmann distributions at different temperatures. These distributions form a 1-d curve parametrized by $T$ in the simplex of normalized distributions.  Whenever the tangent of the quasi-static locus at $\vec \pi(T_b+\Delta T)$ has no contra-variant component\footnote{note that in general the eigenvectors $\vec v_i$ are non-orthogonal, hence the distinction between covariant and contra-variant coefficients is necessary} in the $\vec v_2$ direction  for some $\Delta T>0$, the Mpemba effect occurs for  $T_h>T_c>T_b+\Delta T$.  To show this, we note that any point along the quasi-static locus can be written in the form $\vec{\pi}(T_b+\Delta T) = \vec\pi(T_b) + \sum_i a_i(\Delta T) \vec v_i$. The vector $\vec \alpha(\Delta T)$ given by 
		\begin{eqnarray}
				\vec \alpha (\Delta T) = \sum_i \frac{\partial a_i(\Delta T)}{\partial(\Delta T)} \vec v_i 
		\end{eqnarray} 
		is tangent to the quasi-static locus at $\vec\pi(T_b+\Delta) T$. If for some $\Delta T$ the tangent vector $\vec{\alpha}(\Delta T)$ has no contra-variant component along $\vec v_2$, then $\frac{\partial a_2}{\partial(\Delta T)}$ changes sign at this $\Delta T$ and hence $|a_2|$ decreases with $\Delta T$, and the generalized Mpemba effect occurs. 
		
\begin{figure}
		 	\centering
		 	\includegraphics[scale=0.55]{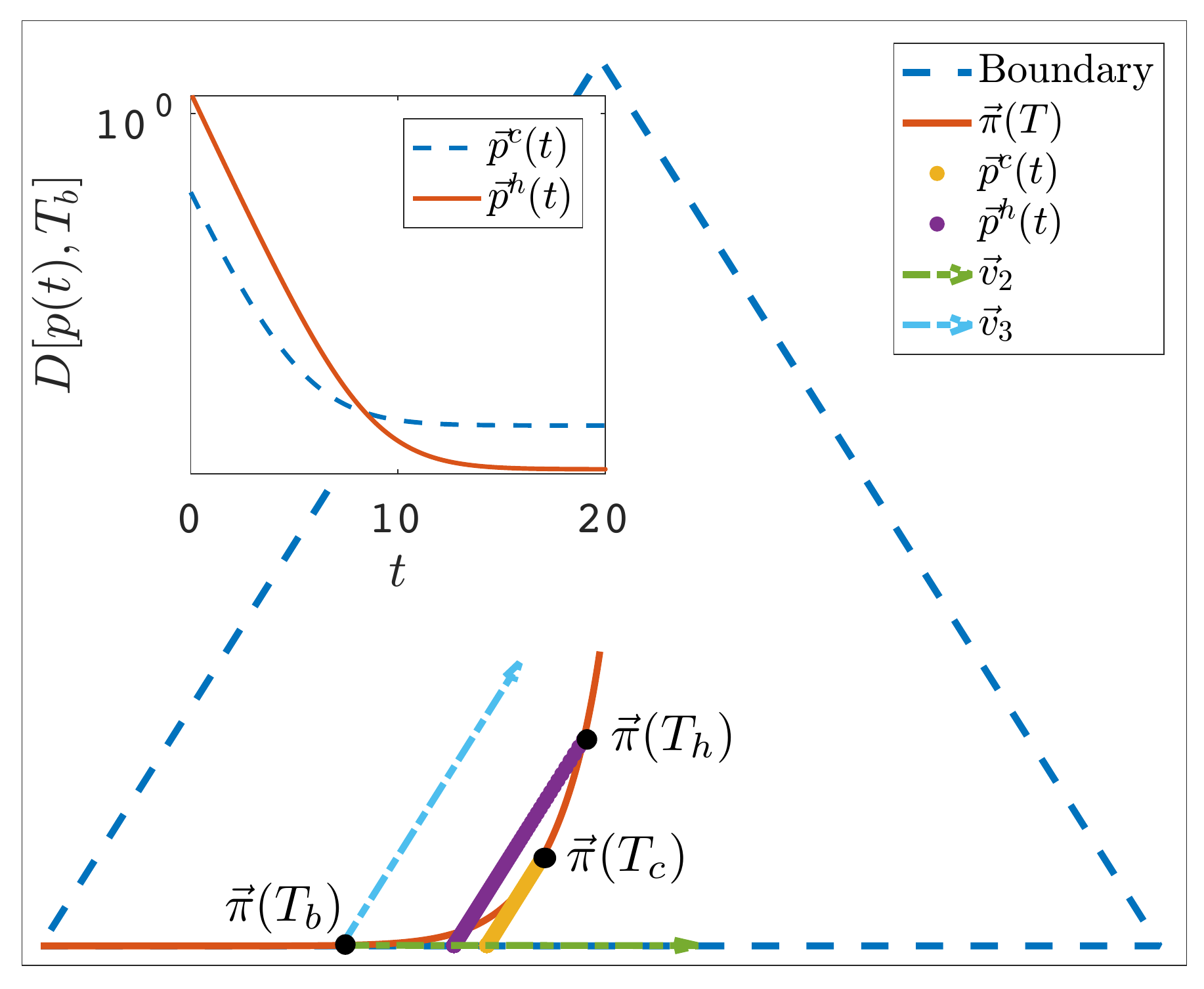}\caption{The set of all normalized probability distributions form a simplex -- the dashed blue triangle. The red solid curve is the quasi-static locus, $\vec \pi(T)$. The dashed arrows are the two right eigenvectors of $R_{ij}$, $\vec v_2$ and $\vec v_3$, associated with fast (blue) and slow (green) relaxation modes. The dotted lines represent the relaxation process, $\vec p^c(t)$ starting at $T_c=0.42$ (orange) and $\vec p^h(t)$ starting at $T_h=1.3$ (purple). The inset shows the distances $D_e[\vec p^h(t);T_b]$ and $D_e[\vec p^c(t);T_b]$, both decrease with time. The initially hot system starts at a larger distance, but after some time its distance from equilibrium is smaller than that of the initially cold system.}
		 	\label{fig:Fig_2_Triangle_with_Examp}
		 \end{figure}

		\paragraph{A minimal three-state model.}	
		 Next we consider an illustrative example of the Mpemba effect in the three-state system shown in Fig. (\ref{Fig:model}a). The energies of the three states are given by $E_1=0$, $E_2=0.1$ and $E_3 = 0.7$, and the energy barriers by $B_{12} = 1.5$, $B_{13} = 0.8$ and $B_{23}=1.2$. The space of all normalized probabilities  in this case contains all the triplets $(p_1,p_2,p_3)$ satisfying $p_1+p_2+p_3=1$ and $0\leq p_i\leq 1$, namely all the points in the 2-d simplex whose vertices are located at $(1,0,0)$, $(0,1,0)$ and $(0,0,1)$ -- see Fig.~(\ref{Fig:model}b). Any initial condition -- a point in the simplex -- evolves according to the master equation toward the equilibrium point, $\vec \pi(T_b)$. The quasi-static locus -- the  curve consisting the Boltzmann distributions with respect to all temperatures $T\in (0,\infty)$ -- is shown as the red-blue solid line in Fig.~(\ref{Fig:model}b).  In non-equilibrium cooling, $\vec p(t)$ typically does not trace the quasi-static locus. For instance, the trajectories initiated at the Boltzmann distributions $\vec p^{~h}(t=0)=\vec \pi(T_h)$ and $\vec p^{~c}(t=0)=\vec\pi(T_c)$ with $T_h=1.3$ and $T_c=0.42$,  are plotted in Fig.~(\ref{fig:Fig_2_Triangle_with_Examp}). These trajectories clearly do not follow the quasi-static locus, though both  $\vec p^{~h}(t)$ and $\vec p^{~c}(t)$ relax toward the same final equilibrium distribution $\vec \pi(T_b)$.  For this system, $D_e[\vec p(t),T_b]$ can be analytically obtained, and is plotted in the inset of Fig.~\ref{fig:Fig_2_Triangle_with_Examp}. After $t_m\approx9$, the distance from equilibrium of the initially hot system drops below that of the initially cold system, $D_e[\vec p^h(t);T_b] < D_e[\vec p^c(t);T_b]$ and hence the Mpemba effect occurs in this case.

	As discussed earlier, a sufficient condition for the Mpemba effect is that there exist a point on the quasi-static locus where the a tangent vector has no contra-variant component along the slowest relaxation direction $\vec v_2$. Here we demonstrate this in the 3-state system. For $ T_b = 0.1$, the fast relaxation vector $\vec v_3$ (blue arrow) is tangent to the quasi-static locus $\{\vec{\pi}(T)|T>0\}$ at $T_{tangent}\approx 0.418$, namely the tangent has no contra-variant component in the slow direction at $T_{tangent}$, and the coefficient in the slow direction decreases beyond this point. Hence, at $T_b=0.1$, \emph{for any two initial temperatures $T_h$, $T_c$ that are larger than $T_{tangent}$ the Mpemba effect can be observed}. This is verified for $T_c = 0.42$ and $T_h=1.3$ (see Fig. (\ref{fig:Fig_2_Triangle_with_Examp})).

\begin{figure}
\centering
\includegraphics[width=0.9\linewidth]{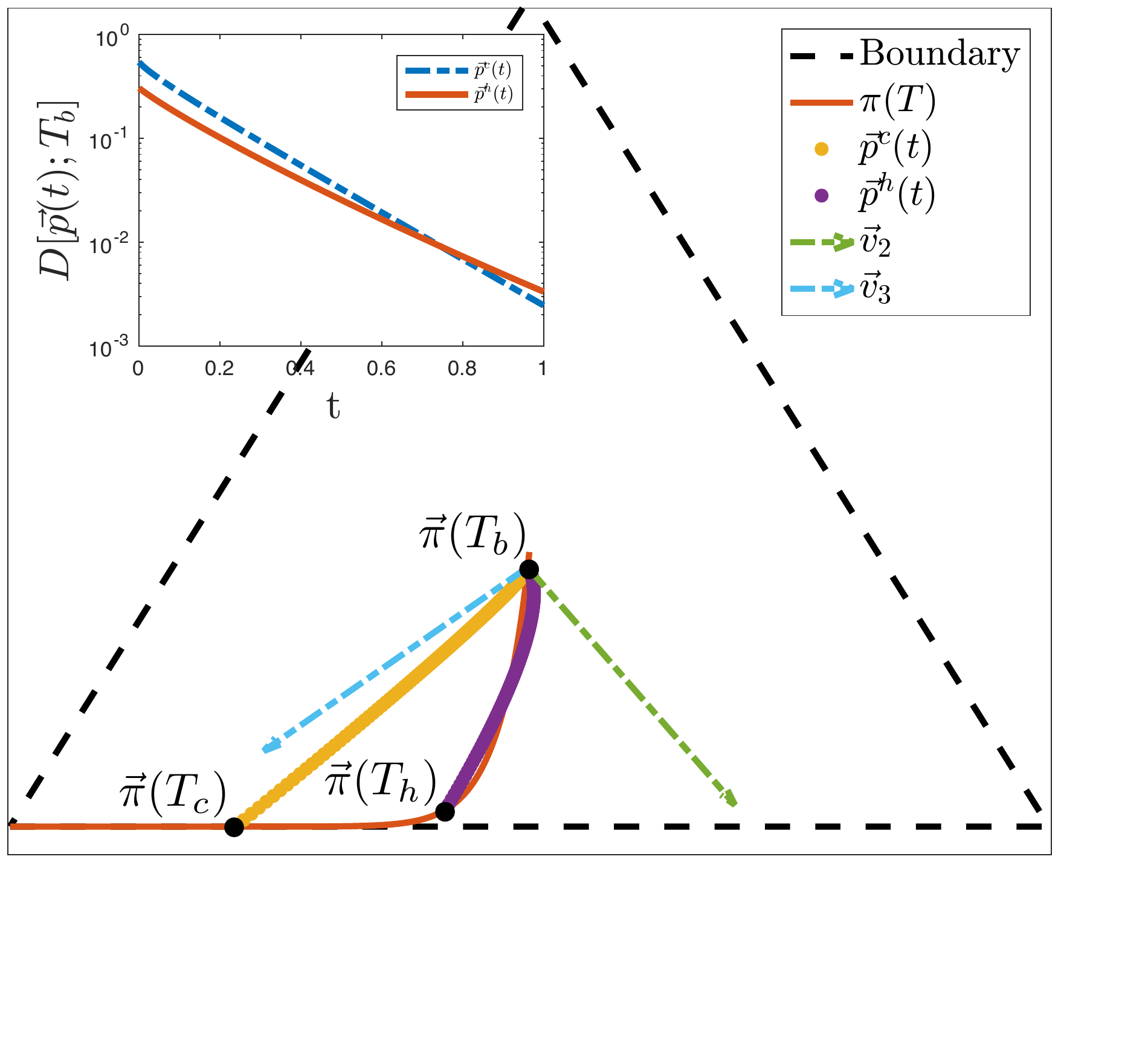}
\caption{
	{\bf The inverse Mpemba effect.} Two identical systems, initiated at a hot temperature $T_c$ and cold temperature $T_h$ are coupled to a hot bath with temperature $T_b>T_h>T_c$. The initially cold system has a very small component in the slow relaxation mode $\vec v_2$ (green dashed arrow), whereas the initially hot system has a much larger coefficient in the slow direction (even though it has a smaller distance from the equilibrium), and hence it decays slower than the initially cold system. Inset -  the distance from equilibrium for both system as a function of time. The initially hot system decays faster.}
\label{fig:InverseEff_DistVsTime}
\end{figure}

\paragraph{The inverse Mpemba effect:} So far we have considered the Mpemba effect for cooling processes. Next, we predict the existence of a similar effect for heating processes, where an initially cold system ($T_c$) heats faster than an initially hot system ($T_h$), when both are heated by the same hot bath ($T_b>T_h>T_c$). 
We illustrate the inverse Mpemba effect by a similar three-state system, with $E_1 = 0$, $E_2= 0.1$, $E_3 = 1$, $B_{12} = 2$, $B_{13} = 1.01$, $B_{23} = 10$. The cold system starts at the equilibrium distribution corresponding to the temperature $T_c = 0.3$, and the hot system at the equilibrium distribution corresponding to $T_h = 0.78$. For both systems, the bath temperature is $T_b = 10$. The corresponding dynamics are shown in Fig.\ref{fig:InverseEff_DistVsTime}, and they clearly show the inverse Mpemba effect.

A sufficient condition for the inverse Mpemba effect is $\lambda_2>\lambda_3$ and $|a_2^h|>|a_2^c|$. This can be seen from essentially the same argument that we used for the Mpemba effect: in this case, $|a_2^h|>|a_2^c|$ implies that for large enough $t$, $\vec p^h(t)$ follows the same trajectory but lags behind $\vec p^c(t)$, and thus the inverse Mpemba effect occurs.

\paragraph{Conclusions} In this manuscript we have discussed a generalized Mpemba effect in cooling and predicted a similar effect in heating (the inverse Mpemba effect). We have found the sufficient condition for a system to have such an effect. Such an anomalous cooling and heating  could be found in a wide range of systems (see discussion in SI). Our analysis indicates that for such systems, a counter-intuitive control of the thermal bath's temperature may be employed to reduce cooling and heating time.


		\paragraph{Acknowledgments:} We thank C. Jarzynski, J. Weeks, S. Deffner, J. Horowitz, R. Remsing, Zhixin Lu  and T. Witten for useful discussions, and Y. Subasi, R. Pugatch and G. Ariel for careful reading. O.R. acknowledges financial support from the James S. McDonnell Foundation. Z.L. acknowledges financial support from the NSF under grant DMR-1206971.


\end{document}